\documentclass[namedreferences]{solarphysics}
\usepackage[hyperref,optionalrh,solaromanenum]{spr-sola-addons} 
\usepackage{graphicx}                    
\usepackage{color}                       
\usepackage{breakurl}                         

\usepackage{siunitx}

\RequirePackage[T1]{fontenc}
\RequirePackage[utf8]{inputenc}

\usepackage{xfrac} 

\newenvironment{disclaimer}[1][Disclosure of Potential Conflicts of Interest]{\footnotesize\paragraph*{#1}}{}

\begin{document}

\begin{article}
\begin{opening}

\title{Neural Network Forecast of the Sunspot Butterfly Diagram}

\author[addressref={add1},corref,email={eurico.covas@mail.com}]{\inits{E.}\fnm{Eurico}~\lnm{Covas}~\orcid{0000-0003-2680-945X}}
\author[addressref={add1},corref,email={nuno.peixinho@gmail.com}]{\inits{N.}\fnm{Nuno}~\lnm{Peixinho}~\orcid{0000-0002-6830-476X}}
\author[addressref={add1,add3},corref,email={jmfernan@mat.uc.pt}]{\inits{J.}\fnm{Jo\~ao}~\lnm{Fernandes}~\orcid{0000-0002-1663-3334}}

\address[id=add1]{CITEUC -- Centre for Earth and Space Science Research of the University of Coimbra, Geophysical and Astronomical Observatory of the University of Coimbra, 3040-004 Coimbra, Portugal}
\address[id=add3]{Department of Mathematics, University of Coimbra, 3001-454, Coimbra, Portugal}

\runningauthor{Covas \textit{et al.}}
\runningtitle{Neural Network Forecast of the Butterfly Diagram}

\begin{abstract}
Using neural networks as a prediction method, we attempt to demonstrate that forecasting of the Sun's sunspot time series can be extended
to the spatial-temporal case. We employ this machine learning methodology
to forecast not only in time but also in space (in this case the latitude) on a spatial-temporal dataset representing the 
solar sunspot diagram extending to a total of 142 years.
The analysis shows that this approach seems to be able to reconstruct the overall qualitative aspects of the spatial-temporal series, namely the overall shape and amplitude of the latitude and time pattern of sunspots. This is, as far as we are aware, the first time neural networks have been used to forecast the Sun's sunspot butterfly diagram, and although the results are limited in the quantitative prediction aspects, it points the way to use the full spatial-temporal series as opposed to just the time series for machine learning approaches to forecasting. Further to that, we use the method to predict that the upcoming Cycle 25 maximum sunspot number will be around $R_{25}=57 \pm 17$. This implies a very weak cycle and that it will be the weakest cycle on record.
\end{abstract}
\end{opening}
\section{Introduction}
\label{introduction}
The sunspots are mostly a visible phenomenon in the solar photosphere and a manifestation of solar activity. One can find extensive bibliography about the solar
cycle, its causes and consequences. Here we emphasise just a few number of previous results and further detail can be obtained in the recent reviews
\citep{butterfly, 2015LRSP...12....4H} and references therein. Since \cite{1908ApJ....28..315H}, it is  known that sunspots contain strong magnetic fields.
These decrease the energy flux considerably and so the sunspots appear darker than the surroundings.

We know, since \cite{1844AN.....21..233S}, that the rise of the sunspots in the solar surface is cyclic but not periodic, with erratic time lapses between maxima and/or
minima that can span between 9 to 13 years. However, one can establish an average period time of about 11 years. Moreover, thanks to the analysis of the
cosmogenic isotopes, \textit{e.g.}\ \textsuperscript{14}C and \textsuperscript{10}Be \citep{1998SoPh..181..237B} it is possible to reconstruct the solar cycle back to more
than \SI{10000} years which is particularly interesting for paleoclimatology \citep{2004Natur.431.1084S}. Recently, \cite{2017Geo....45..279L} have shown some evidence that
there is a cycle around 11 years on 300-million-year fossilised tree rings. This could mean that the sunspot cycle has been around for
a much longer time scale than our current direct observation record.

Being weakly chaotic \citep{1988ssgv.conf...69W, 1990RSPTA.330..617W, 1991JGR....96.1705M, 2006A&A...449..379L, 2009SSRv..144...25S,2014SSRv..186..525A} and one of the longest continuously recorded daily measurement made in science \citep{2013Natur.495..300O}, the sunspot series is rightly considered as one of the top benchmarks for time series forecasting.

There is quite a large body of research on forecasting the sunspot time series, in particular the strength, the 
length and the maximum of the next cycle. However,
as indicated by the analysis presented at the American Geophysical Union (AGU) 2008 meeting 
\citep{pianoplot,2012SoPh..281..507P}, there seems to exist more forecasts than possible future data scenarios, as shown in the ``piano plot''
 introduced by William Dean Pesnell, in his summary of the literature for predictions of the current cycle maximum solar spot number. Depending on the most realistic maximum number of sunspots \citep{2017ApJ...839...98A},
one would think that there are more articles predicting the next cycle sunspot maximum than attainable physical possibilities.
Pesnell's plot emphasises that metrics such as the sunspot number 
may not contain enough information to decide among distinct forecasting methods. 

It was under these motivations that one
of us \citep{covas2016} attempted to use a model based on spatial-temporal embeddings to forecast the sunspot diagram,
  demonstrating how a pure mathematical model can be used to do a spatial-temporal forecast (latitude and time), based on the full dataset for the solar spot coverage area and its corresponding latitude, from
1874 up to 2015. 
That work, however, was not the first one to attempt to predict or reconstruct the entire sunspot diagram (space and time, not just time).
\cite{2011A&A...528A..82J} analysed the sunspot diagram in its entirety to calculate correlations between several quantities of sunspot groups against the cycle strength and phase. With these,
they were then able to reasonably reconstruct the sunspot spatial-temporal pattern. 

In this article, we
propose an approach by forecasting the full spatial-temporal sunspot diagram, using exactly the same data as in \cite{covas2016}, 
by means of neural network techniques. A very large number of authors (from the early 1990's up to now) have already
attempted to use neural networks to forecast pure temporal aspects of the sunspot cycle \citep[see references in these reviews]{
2012SoPh..281..507P,Pesnell_2016}.
All of those authors, however, have limited themselves to forecasts in time only, not space and time.
So, in contrast and as proposed, we attempt to forecast the full sunspot diagram, by means of neural networks.
We believe that by including one further dimension to the usual one-dimensional forecasts in the literature,
one should, in the future, be able to better decide between the multitude of forecasting methods.

The article is organised as follows. In Section \ref{method} we introduce the method using neural networks for forecasts and give details of how to apply the approach to a dataset with one spatial and one temporal dimension. 
 In Section \ref{results} we apply the approach to the sunspot dataset. Finally, in Section
\ref{conclusion} we draw our conclusions and suggest future research ideas.

\section{The Method: Feed-Forward Neural Networks}
\label{method}


Here we propose an approach which is based on artificial neural networks  
\citep{Lecun98gradient-basedlearning, Krizhevsky_imagenetclassification}. We use a subset of neural networks
called feed-forward artificial neural networks. These are basic networks that have
an input layer, one or several hidden layers, and an output layer, fully connected
but where no connection occurs backwards or on a loop.
We train the network using the back-propagation algorithm \citep{david_mcclelland_jamesrumelhart1989, 1986Natur.323..533R} and
we use on-line training \citep[see][and references therein]{9780262527019} as recommended in the 
 literature \citep{Wilson20031429}.

Most of the neural network literature on forecasting can be divided into two groups, one using time delays \citep{Kim199948} 
to construct the vector input patterns
for the feed-forward neural network and another using what is called a recurrent network \citep{COGS:COGS203}, where the information is allowed to cycle in a loop
\citep{petrosian2000recurrent, zhang2000predicting, han2004prediction, 9783838303826, Chandra:2012:CCE:2181341.2181747}. Here we shall use only the former, mainly for simplicity, easiness of implementation, and interpretability. 
The time delay approach in the literature uses the embedding dimension and sometimes the
derived time lags to create an appropriate set of input patterns 
\citep[see, \textit{e.g.}][and references therein]{
GEAN:GEAN12026}. 
The extension to spatial-temporal forecasting follows the same approach, \textit{i.e.}\ using lags and an embedding dimension, but in space and in time.
The approach results from merging this technique with another forecasting method unrelated to neural networks and introduced in
\cite{2000PhRvL..84.1890P} for the
reconstruction of spatial-temporal datasets. In their article they used their approach successfully to both
a spatial-temporal version of the H\'enon map and to a synthetic dataset arising from evolving the Kuramoto-Sivashinsky non-linear model. 
Following on \cite{2000PhRvL..84.1890P}, there was later an attempt to forecast financial spatial-temporal datasets by \cite{Covas}, with reasonable success, and also an attempt to use it for the  sunspot  diagram by \cite{covas2016}. 
In those articles, a grid of inputs was built based on the two dimensional
data series $s(n,m)$ to use in the search of the nearest neighbour in the embedding space, that once found, gave an approximation to the true future evolution
in the original space. 
We shall describe this approach in Section \ref{Parlitz-Merkwirthmethod}. 

Although there is some limited research on the application of neural networks to spatial-temporal forecasting in several distinct settings \citep[see, \textit{e.g.}][and references therein]{
2017arXiv170805094M,2018Chaos..28d1101P, 2017Chaos..27l1102P,
2017Chaos..27d1102L,PhysRevLett.120.024102,2017arXiv171110566R,2017arXiv171110561R, 2018arXiv180106637R,
2018JCoPh.357..125R}, we believe that this is the first time that neural networks have been  applied in the specific context of  sunspot spatial-temporal forecasting.

\subsection{Input Layer Architecture}
\label{Parlitz-Merkwirthmethod}

The input layer design we take can be seen as a spatial-temporal generalization of the 
time delay neural network method \citep{Waibel:1990:PRU:108235.108263, luk2000study, Frank2001, OH2002249, 1009-1963-12-6-304} together with a merge of the spatial-temporal method of Parlitz-Merwirth \citep{2000PhRvL..84.1890P}, applied to feed-forward neural networks.

\begin{figure*}
\centering
\resizebox{\hsize}{!}{\includegraphics{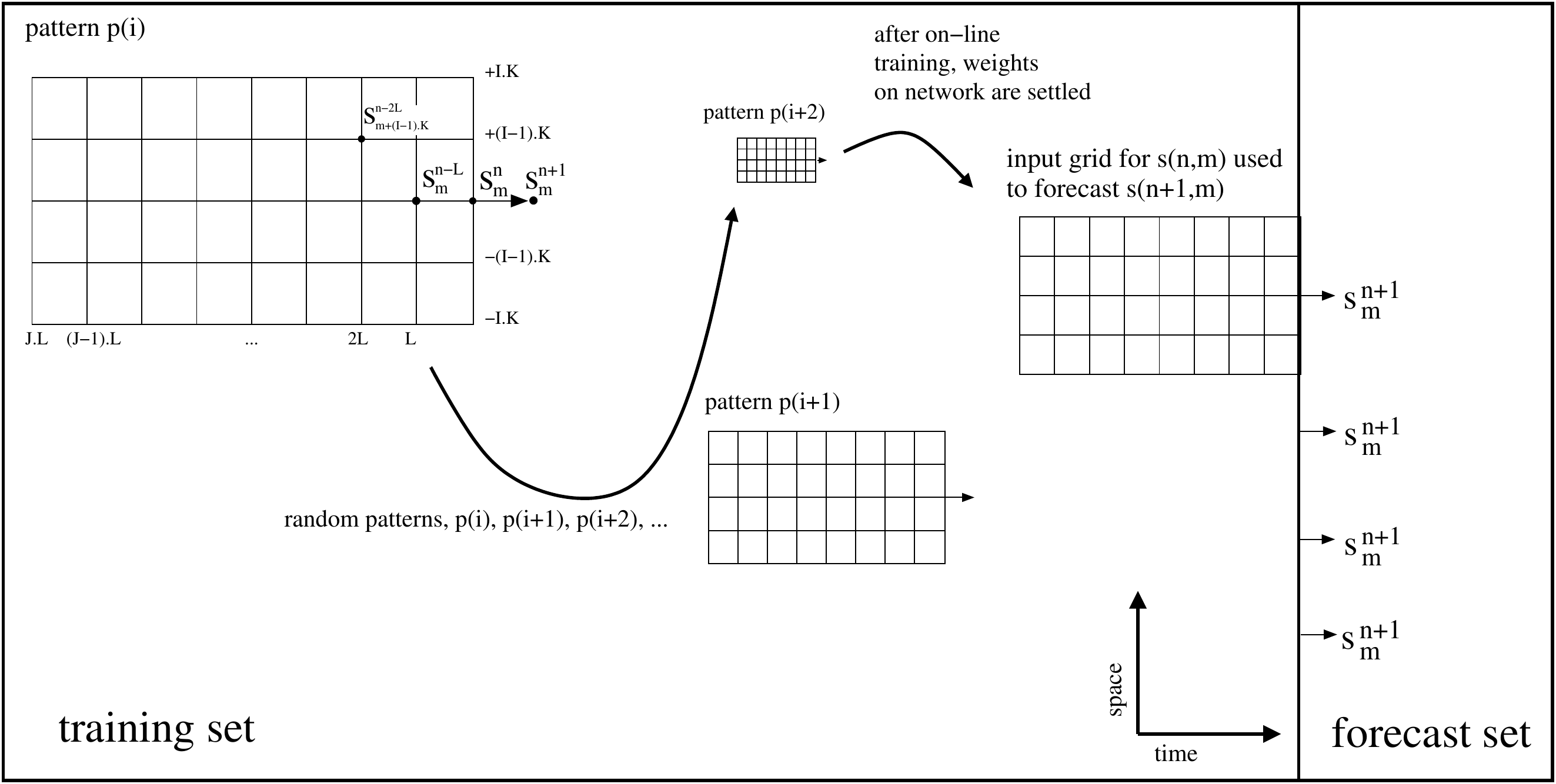}}
\caption{Forecasting method illustration. One constructs an embedding space using delays, then assembles randomly positioned grid input patterns within the training set to pass to the neural network (in this figure
we show 3 randomly selected input patterns).
The input is a
$(2 I+1)(J+1)$ vector ${\bf x}(s^n_m)$ and the target (output) to train the network is the value $s^{n+1}_{m}$.
After training with a sequence
of patterns $p(i), p(i+1), p(i+2), \ldots$ then the patterns adjacent to the forecast set are used to calculate the outputs
to compare against the forecast. To forecast the $n+2$ slice we concatenate the previously predicted $n+1$ and progress accordingly.}
\label{NeuralNetwork}
\end{figure*}

The model is constructed using the concept of super-state vector ${\bf x} (s^n_m)$ as described already in great detail
 \citep{covas2016} and we refer the reader to it for further reading. These super-state vectors have the following
formula:
\begin{eqnarray}
\label{embedding}
{\bf x}(s^n_m)=&\{&s^n_{m-I K }, \ldots,s^n_m, \ldots, s^n_{m+I K},\\
               & &s^{n-L}_{m-I K},\ldots, s^{n-L}_{m},\ldots, s^{n-L}_{m+I K},\ldots \nonumber\\ 
               & &\ldots,s^{n-J L}_{m-I K},\ldots, s^{n-J L}_{m},\ldots, s^{n-J L}_{m+I K} \},
\nonumber
\end{eqnarray}
and are built from a rectangular grid of the original spatial-temporal data series ${\bf x} (s^n_m)$. Again, as in  \cite{covas2016}, $I$
refers to the number of neighbours on each side in space, $J$ to the number of temporal ones, and $K$ is the spatial lag, while $L$, the final parameter, is the temporal lag.
The $K$ and $L$ parameters, are chosen again as in \cite{covas2016}. In that article the 
average mutual information method \citep{Fraser86, abarbanel1997analysis} was used  to calculate the optimal spatial-temporal delays,
and we use the same (we refer the reader to that article for further details). Regarding the embedding parameters $I$ and $J$, again we follow
the approach as in \cite{covas2016} and use the false nearest neighbours algorithm \citep{1992PhRvA..45.3403K}. These two approaches
allow the feature selection architecture to be defined automatically before we attempt any neural network forecast.

These input vectors ${\bf x}(s^n_m)$ (see Figure \ref{NeuralNetwork}), 
 are then passed as input to the neural network grid while the target (output) is $s^{n+1}_m$.
Using this method, the number of input neurons is the same as the dimension of ${\bf x}(s^n_m)$, \textit{i.e.}\ $(2I+1)(J+1)$. 
It seems reasonable to define the input layer architecture in this way, and after extensive searches, we find that this architecture, with its four auto-calibrated parameters ($I$, $J$, $K$, $L$), seems to be optimal for forecasting. 
The number of output neurons needed for this approach
is just one, \textit{i.e.}\ the predicted $s^{n+1}_m$. 
Finally, the number of hidden nodes 
will be decided by trial and error later on, given that there seems to be no theoretically firm formula for the optimal number
of these.

\section{Results}
\label{results}

\subsection{The Dataset}

The dataset used is available freely in \cite{butterfly} and is exactly the same used in \cite{covas2016}, in part to make sure we can do a comparison between the model presented here,
\textit{i.e.}\ a neural network forecast, against the model in \cite{covas2016}, which was a non-linear embedding model. We refer the reader
to that article for finer details of the dataset. To summarize here, the data is a latitude and time set of sunspot areas, covering 
142 years. The base training set used the first 1646 temporal latitudinal values and the base test set is made of the remaining 242 time latitudinal values.

\subsection{Neural Network Parameter Calibration}
\label{parameters}

Given that we use exactly the same base dataset as in a previous article \citep{covas2016}, we can use exactly the same calibration for the feature selection architecture parameters, \textit{i.e.}\ $I$, $J$, $K$, $L$ in Equation \ref{embedding}. 
The calibrated parameters for this base training set were $I=2$, $J=6$, $K=9$, and $L=70$, \textit{i.e.}\  each
${\bf x}(s^n_m)$ is made of $(2I+1)(J+1)=2\times2+1=5$ spatial neighbours by $6+1=7$ time delayed neighbours, with a spatial lag of $9$ latitudinal slices and 
a temporal lag or delay of $70$ Carrington rotations or time slices, corresponding to approximately 5.22 years. We shall use these choices of parameters
 to create training sets
for the neural network and after the training, the input sets for the forecasts as depicted in Figure \ref{NeuralNetwork}. Notice that this feature selection or feature representation choice,
as done in \cite{covas2016}, only depends on the training set, and therefore up to here, this approach is self-contained and auto-calibrated.
However, other neural network parameters have to be decided empirically, as there is no agreed way to calculate them \citep[see][and references within]{stathakis2009many}. 
The most important is the number of hidden neurons on each hidden layer and the number of those layers. 
To keep the approach as simple as possible, we decided to use only one hidden layer. 
Regarding the number of hidden nodes $N_h$, there is again, as far as we know, no universally agreed procedure to decide what is the optimal number. Using too
little will result in under-fitting, and too many in over-fitting, or worst, in fitting the noise of the input data patterns. 
In our case, we have calculated the optimal number of hidden nodes by trial and error, as our particular case study does not require large
computational power (an entire run with one million iterations takes a few minutes to complete on an average personal computer). 
We have seen that the neural network
can produce realistic results when the number of nodes is between 50 and 100 or so, with an optimal value of $N_h=70$ nodes. We shall use this value hereafter. 
We note that there are some algorithms, namely
``pruning'' and ``constructive algorithms'' \citep{Cun:1990:OBD:109230.109298, Hassibi, 9780262527019} that try to overcome this problem,
by starting with a large network and calculating which of the weights or links on the network 
are superfluous. For the purpose
of this article, given that the dataset is small, 
we will keep it as simple as possible and stay away from pruning and heuristic approaches and just do an exhaustive search.
As for the back-propagation hyper-parameters, $\eta$ and $\alpha$, representing respectively the rate of learning and momentum of the algorithm, we shall use, based on our searches, the values of $\eta=0.3$ and $\alpha=0.01$. Notice that we
use a variation of the algorithm with an adaptive learning rate $\eta_n=\sfrac{\eta}{\left(1+n/10000\right)}$, where $\eta_n$ is the learning rate used at time step $n$.
In addition to the depth and width of the hidden layer(s) architecture, another important degree
 of freedom is the choice of the activation function. 
 Here for simplicity, we shall use the logistic or sigmoid function $\sfrac{1}{\left(1 + e^{-x}\right)}$
as the activation function for all layers.
Another parametrization is what normalization to take \citep{9780262527019}.
In the case of the sunspots, we have area values from a minimum of 0
to a maximum of \SI{2580} in units of millionths of a hemisphere. One approach is to subtract the mean and divide by the standard deviation.
However, the sunspot area distribution function does not follow a Gaussian distribution. In fact, the distribution is closer to
a power law with an exponential cut-off, similar to the probability distribution function of the so called in-out intermittency
\citep{1999Nonli..12..563A}. So, we take another approach, namely to apply the transformation:
$$
x \to \alpha_{nor}+\frac{\ln(1+x)}{\beta_{nor}},
$$
where $\alpha_{nor}$ and $\beta_{nor}$ are the arbitrary shift and scaling constants, respectively. Again, there is no clear rule on how to choose these and
we calibrate them by trial and error. We have seen that the neural network
can produce realistic results when the $\alpha_{nor} = 0$ and $\beta_{nor} = 10$. We shall use these values hereafter.
Our final free set of parameters relates to the initialization of the weights. We choose random numbers with a constant distribution between $[0,1]$ and shifted by $\alpha_{rng}$ and scaled
by $\beta_{rng}$. We find that we obtain good results (as measured by the similarity index introduced below) around $\alpha_{rng} = -0.5 $ and $\beta_{rng} = 0.01$.
We notice that for all parameters above, we conducted extensive stress testing and chose the parameters values above mentioned for which the similarity between the spatial-temporal forecast and the original real data was maximal.

\subsection{Training and Forecast}
\label{originalforecast}

To train the neural network, we have used one million iterations of different input patterns. Notice that the maximum number of different input patterns is 
$1646\times50=$ \SI{82300} patterns. However, for neural network training, one randomly chooses patterns to optimize the weights, even if the number of
iterations exceeds the number of unique patterns,
and using this approach is equivalent to a stochastic progress towards the minima of the error function \citep{9780262527019}. 


\begin{figure}
\resizebox{\hsize}{!}{\includegraphics{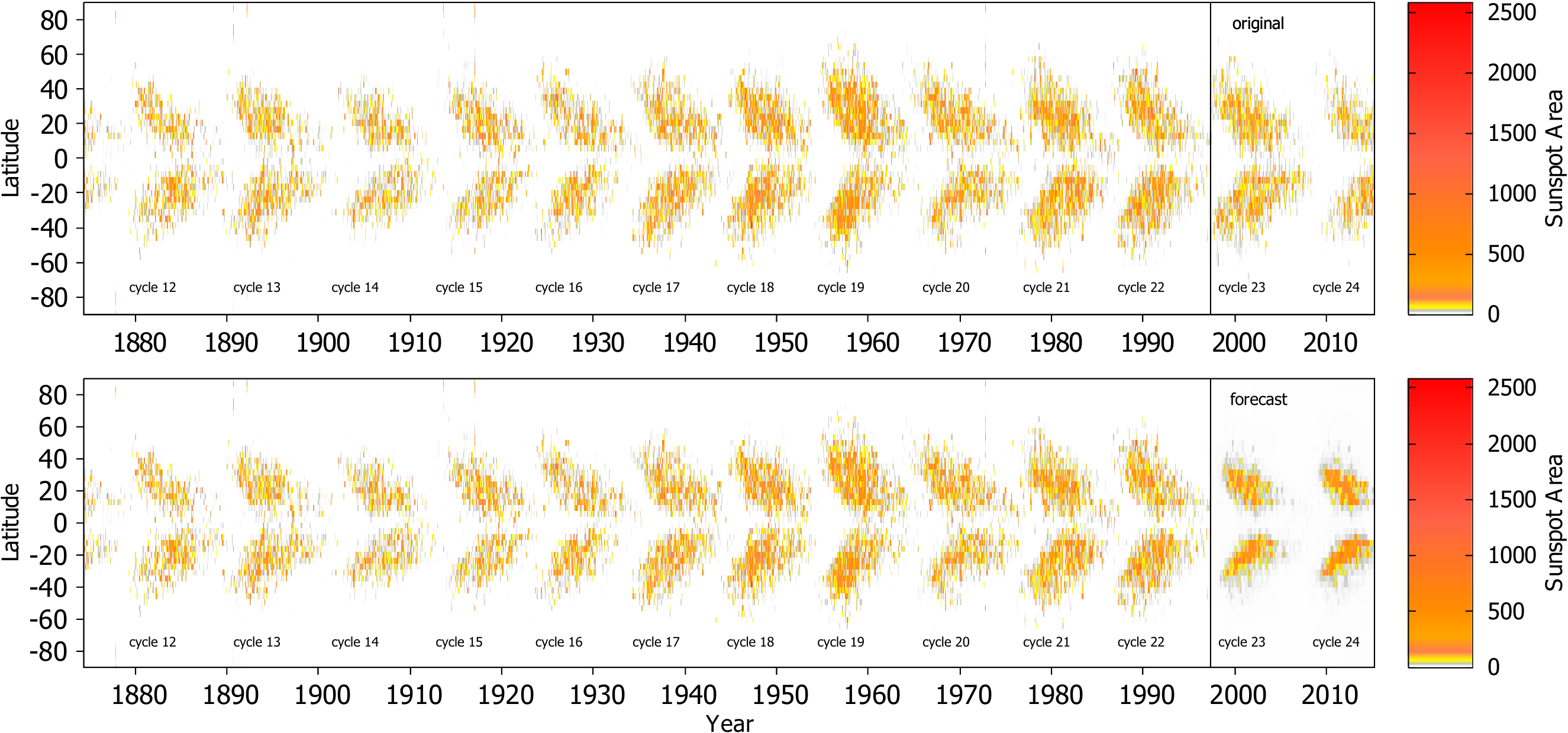}}
\caption{Forecast of the last two sunspot solar cycles using spatial-temporal neural networks. We have used the parameters $I=2$, $J=6$, $K=9$,
 and $L=70$
(as in \citealp{covas2016}). This figure is very similar to Figure 6 in that article, using exactly the same original dataset, but while \cite{covas2016} used a non-linear embedding method, we used the neural network method for the forecast. The fact that both methods result in similar
forecasts and corresponding figures is reassuring, as it reinforces the adequacy of the feature selection approach based on non-linear embeddings.
The specific neural network hyper-parameters used were:  $\eta=0.3$, $\alpha=0.01$, $\alpha_{nor} = 0$, $\beta_{nor} = 10$, $\alpha_{rng} = -0.5 $, $\beta_{rng} = 0.01$, $N_h=70$ hidden nodes, run  
with one million pattern iterations, and the 
 logistic or sigmoid function as the activation functions on both the hidden and output layer. The upper panel represents the sunspot dataset, divided into the base training set and the base test set. The lower panel represents the training together with the respective
 forecast. On both panels the split between training set and observed future dataset or forecast set is shown with a black line. 
  Notice that, as in \cite{covas2016}, the main
characteristics of the sunspot diagram are reproduced, namely the overall intensity of the cycle and the sunspot band progress towards the equator are both present. Overall, the forecast is quite good from the qualitative point of view.}
\label{Spatiotemporal_Forecasting_Sunspots_LastCycle}
\end{figure}

\begin{figure}
\resizebox{.9\hsize}{!}{\includegraphics{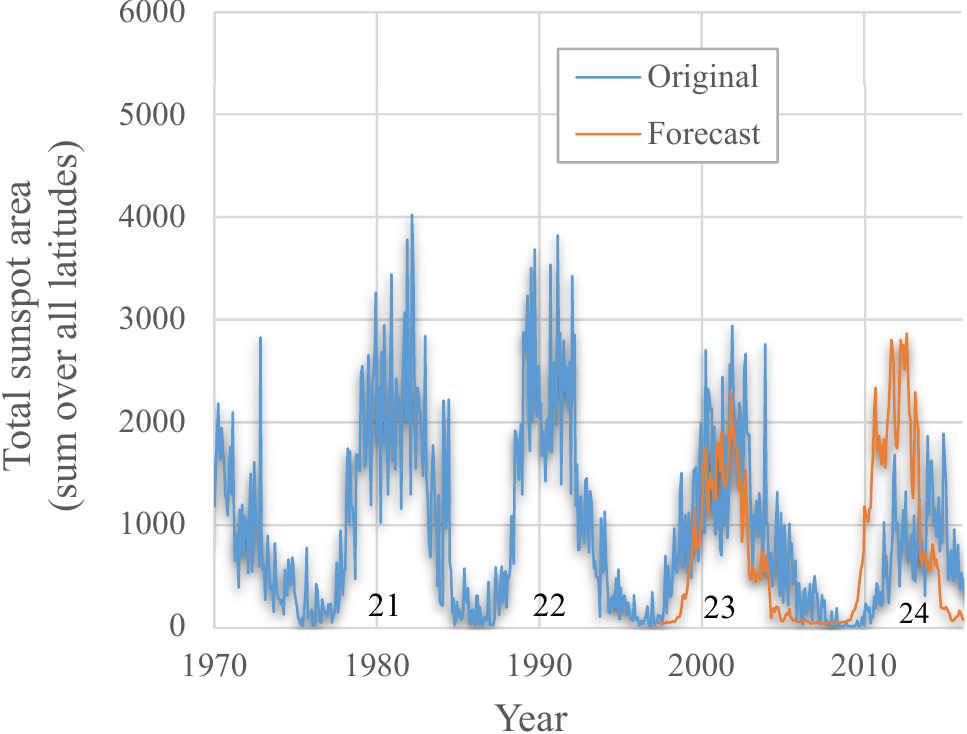}}
\caption{Latitudinal sum of the sunspot area $A(t)$ and the forecast for the two last Cycles (23 and 24) - same scenario
as in Figure \ref{Spatiotemporal_Forecasting_Sunspots_LastCycle}. We have marked the cycle numbers near the horizontal axis. The forecast for Cycle 23 is reasonable but the method fails badly for Cycle 24, showing clearly the limitations of the approach
at least in its current format.}
\label{SolarForecastingNeuralNetworksSumOverLatitude}
\end{figure}

Using the best parameter calibrations from Section \ref{parameters}, we can then proceed to forecast the base test set (made of Cycle 23 and the beginning of Cycle 24). We choose to split the full dataset into a training set consisting of data up and including Cycle 22 inclusive,
with the remaining data being the forecast set. We choose this split partly because we want
to forecast at least one entire cycle, and partly because we want to do a consistent comparison with the previous results in \cite{covas2016}.
The model outputs one latitudinal vector at a time, and we then stack that vector to the training set, and reuse the same 
calibrated
parameters to forecast the next latitudinal vector. The results of this forecasting are depicted in
Figure \ref{Spatiotemporal_Forecasting_Sunspots_LastCycle}.
The results show that this method can reproduce the two main
features of the sunspot spatial-temporal original series, \textit{i.e.}\ the amplitude shape of the cycle (resembling a butterfly wing, hence the usual name given to the sunspot cycle) and the sunspot band or wing progresses or moves towards the equator, although there are some quantitative differences, \textit{e.g.}\ there seems to be a concentration of points on the forecast and the butterfly wings look denser.
To 
strengthen this conclusion, we also depicted in Figure \ref{SolarForecastingNeuralNetworksSumOverLatitude} the calculation of the total sunspot area (summed over latitudes) for the forecast and compared it with the
observed one. It shows the total sunspot area time series, from both the original training set
and the forecast set, and even if both sets are noisy, it demonstrates
that the neural network approach can work not only in space and time but also on one-dimensional reductions like the latitudinal sum.
This result seems to indicate that the neural network method is reasonably good at forecasting
the first cycle but struggles when forecasting ahead to the next cycle. This is however not surprising, as usually the solar sunspot series is recognized as a seminal case of low dimensional chaos. It implies that there is a temporal limit for any reasonable predictability. 

\begin{figure}
\resizebox{.9\hsize}{!}{\includegraphics{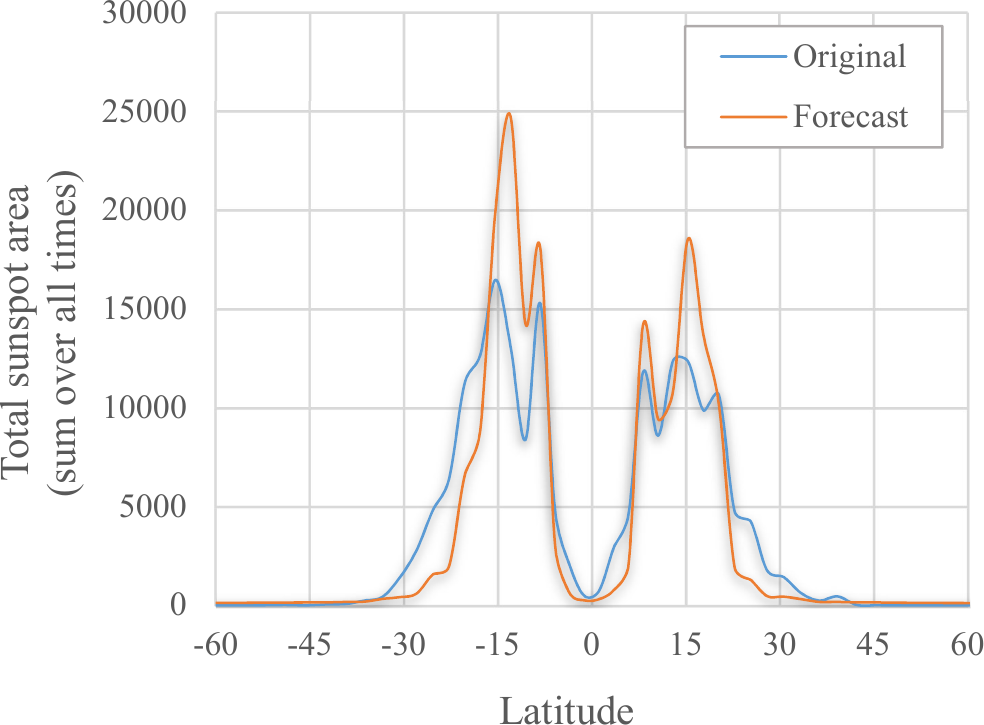}}
\caption{Temporal sum of sunspot area $A(\theta)$ and the forecast for the two last cycles - same scenario as in Figure \ref{Spatiotemporal_Forecasting_Sunspots_LastCycle}. The forecast gives a very similar latitudinal profile, with the
two sunspot bands at mid-latitudes shown clearly on the plot.}
\label{SolarForecastingNeuralNetworksSumOverTime}
\end{figure}

Further to the temporal average, we can also calculate the sum over time, to show the average sunspot area as a function of latitude, another aggregated
metric. This is shown in Figure \ref{SolarForecastingNeuralNetworksSumOverTime} and demonstrates that the method seems to be able to reproduce the real features
 in 
latitude as well as in time. 


\subsection{Structural Similarity}
\label{ssim}

The novelty of the sunspot butterfly diagram forecasting approach presented here is that it attempts to predict in both space and time using neural networks. The question is, how does one verify that a forecast is good quantitatively as opposed to qualitatively?
As in \cite{covas2016}, we employ a widely used computer vision metric called structural similarity index, $\textnormal{SSIM}(x,y)$, which was first 
introduced in \cite{Wang04imagequality}.
The SSIM index is an numerical quantity commonly used to calculate the discerned quality of images and videos.   A value of $\textnormal{SSIM}=1$ corresponds to the case of two
perfectly identical datasets or images, in our case here, a perfect spatial-temporal forecast.
We shall use now the SSIM metric to calculate the similarity of the forecast and the original sunspot cycle. We verified that, as we randomly shift our neural network free parameters, for similar looking forecasts (as
quantified by the human eye and the SSIM index), the root mean square error (RMSE) value could oscillate widely.
The underlying reason for the failure of the RMSE metric is that the data is not a continuous variable, but a very irregular (or spiky) dataset. This is why we shall use the SSIM as our goodness of forecast metric. 

We should also note that we should attempt to forecast other cycles besides Cycle 23 and Cycle 24, as  
the actual effectiveness of the method in terms of actual predictability may vary cycle to cycle, as already shown for the non-linear embedding method in \cite{covas2016} and 
because the overall details of each cycle are quite variable \citep[see \textit{e.g.}][]{
2003ApJ...589..665H, 2011SoPh..268..231I}.
The method depends only on the training set, and if a cycle is relatively weak, then unless we have enough training set examples of embedding state vectors corresponding to weak cycles, the forecast will surely fail. The same 
presumably would apply for 
strong cycles. 
To address this question, we calculated the SSIM index for all the cycles one by one against a training set made of all the data from the beginning up to that cycle, using the information that can be found in Tables 1 and 2 in \cite{2015LRSP...12....4H} that gives approximate dates of cycle 
beginnings and ends.
\begin{figure}
\resizebox{\hsize}{!}{\includegraphics{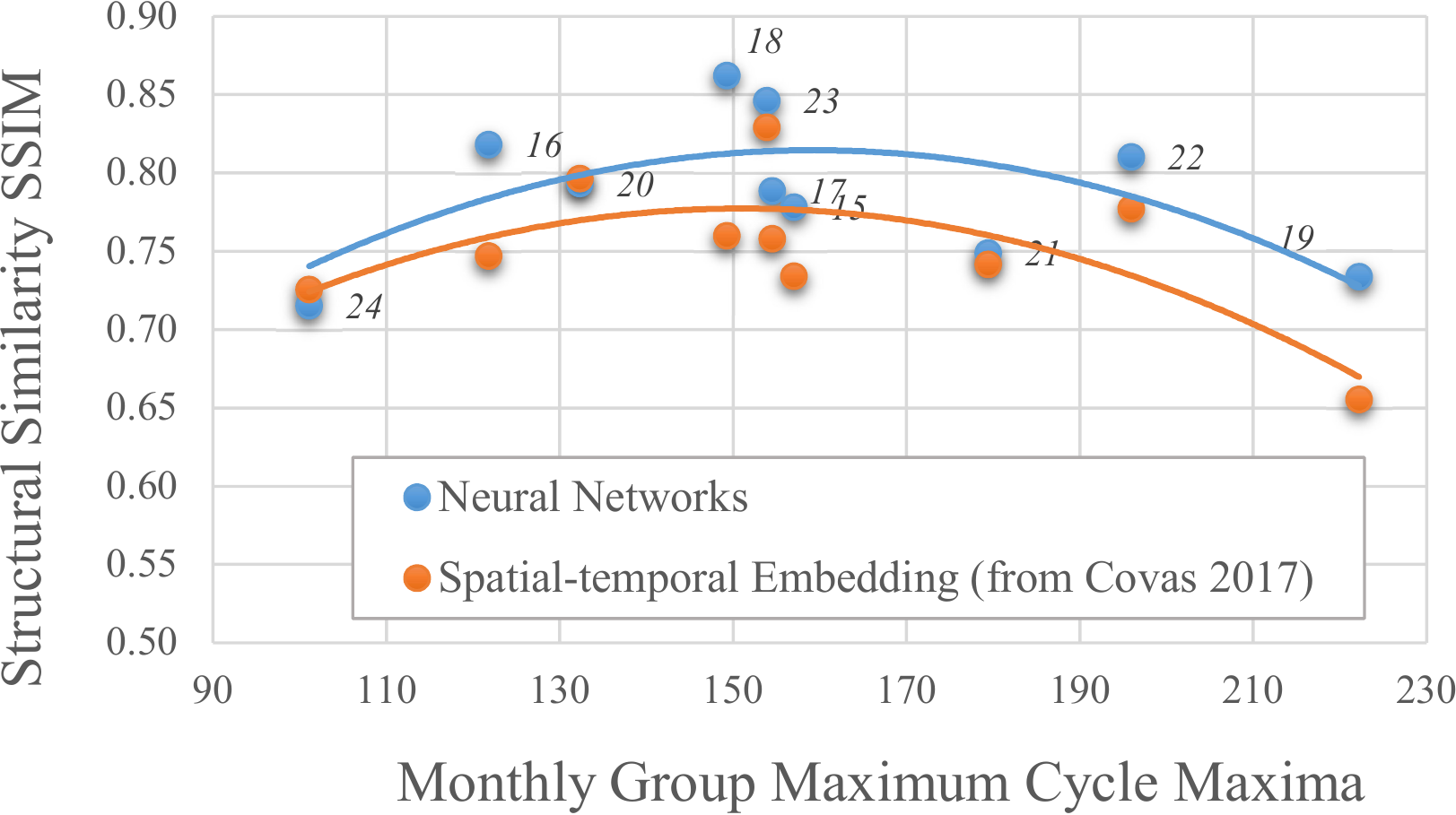}}
\caption{The structural similarity, SSIM, of the prediction set \textit{versus} the original test set
 against the cycle strength as given by the monthly group maximum sunspot number \citep{2015LRSP...12....4H}.  The plot labels represent the cycle number. The analysis shows that the approach
is most accurate for medium amplitude cycles. Notice we have excluded attempts to predict Cycles 11 to 14 as for these cycles there is not enough past data \citep[see][for details]{covas2016}. 
To increase the precision of the forecast, we took 20 million patterns \textit{per} cycle, to compensate for the 
smaller average learning set lengths than in the previous figures. 
We compared our results against those in \cite{covas2016}, who used a spatial-temporal non-linear embedding approach. There is a clear similarity in both methods, both being less effective for weak and strong cycles. Overall, we observe that the neural network approach has a small advantage over the 
spatial-temporal non-linear embedding approach.}
\label{SSIM_versus_cycle_maxima}
\end{figure}
The results of our analysis are depicted in Figure \ref{SSIM_versus_cycle_maxima}. There are three conclusions from analysing it.

First, the forecasting ability of the neural network increases as the number of temporal slices increases, and as a result SSIM metric improves a bit. This is a well known behaviour of neural networks, which are very data hungry, \textit{i.e.}\ the larger the number of training patterns, the closer
will the neural network internal parameters (\textit{i.e.}\ the weights) be to the optimal theoretical values.

Second, and most importantly, the neural network approach achieves its higher performance for medium cycles, \textit{i.e.}\ the most common cycles. We have, as in \cite{covas2016},
 fitted a parabola trend on it, that shows this clearly. We also compared and super-imposed the results in \cite{covas2016}, who have used a non-linear embedding method. It seems to show that the neural network method is somehow slightly better at forecasting than the embedding method. However,
 neither of the methods is good enough as one could wish for weak or strong cycles. 
The neural network approach works on the basis of extracting knowledge from the training patterns. If there are not enough
training examples tracing weak or strong cycles, then the weights will converge to values that tend to be biased towards medium cycles - the most
common ones. There are two obvious ways, while using neural networks, to improve the accuracy of the forecasts. One is to gather more sunspot butterfly diagram data. This is not easy, as one either has to physically wait to grow the dataset, and given the average cycle has a rough periodicity of 11 years, one could be waiting a long time. Another way to collect more data is to use recently recovered sunspot butterfly diagram data going back to the early 18th century (see Figure 2 in \citealp{2009SoPh..255..143A} and 
Figure 1 in \citealp{2009ApJ...700L.154U}). The other way to improve accuracy is to include extra data, or as commonly known in machine learning,
extra features. There is a strong argument on using other datasets, such as geomagnetic data \citep{2009ApJ...694L..11W},
solar magnetic fields datasets 
\citep{
2013PhRvL.111d1106M}, the so-called polar faculae \citep{2012ApJ...753..146M},
and solar seismological data \citep{
2013ApJ...777..138I}. Nonetheless,
it is outside the scope of this article to include further data. We note, however, that the neural network method allows the inclusion
of other data easily as long as the temporal frequency of data points is
the same as the sunspot data. We plan to pursue this research line in a forthcoming article. Here we have basically attempted to demonstrate
the possibility of qualitatively forecasting using a pure mathematical method, that is, another approach to be added to the existing ones in the literature, 
such as the application of empirical relationships
\citep{2011A&A...528A..82J, 2015A&A...580A..62S, 2016ApJ...823L..22C} or 
 the ones using solar surface magnetic field datasets \citep{2014ApJ...784L..32M,2014ApJ...792...12M,2017arXiv170700268J}.

Third and finally, Figure \ref{SSIM_versus_cycle_maxima} shows that the neural network method performs slightly better than the non-linear
spatial-temporal embedding approach in \cite{covas2016}. This is quite encouraging that this work is going in the right direction.

\subsection{Further Forecasts}
\label{furtherforecasts}

\begin{figure}
\resizebox{.9\hsize}{!}{\includegraphics{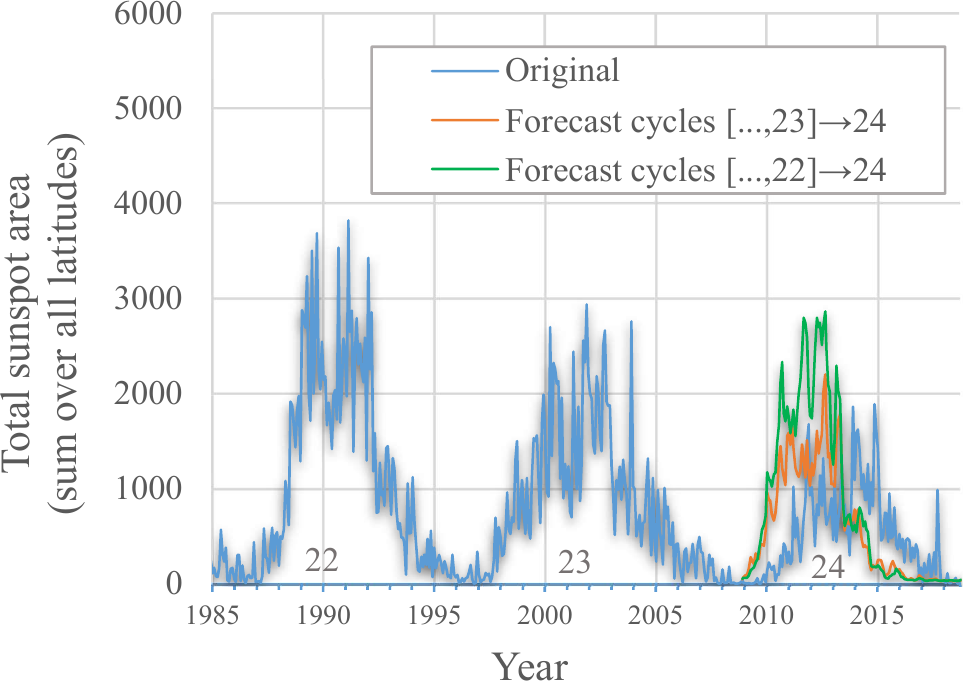}}
\caption{Comparison of the latitudinal sum of the sunspot area $A(t)$ and the forecast for the current Cycle 24 using a training set that includes all the data up to Cycle 23 against the one as in Figure  \ref{SolarForecastingNeuralNetworksSumOverLatitude}. We have marked the cycle numbers near the horizontal axis. It seems to show that using more data improves the forecast, if not in terms of the onset of the cycle, at least in terms of the cycle maximum amplitude.}
\label{SolarForecastingNeuralNetworksSumOverLatitudeCycle24}
\end{figure}

In this section, we examine further forecasts, outside our base training and testing set split we used above. In particular we wanted to understand
the forecast of the most recent Cycle 24 and the future Cycle 25. We have seen from Figure \ref{SSIM_versus_cycle_maxima} that the forecast of Cycle 24, a weak cycle, was not very good in terms of the SSIM index. In order to know if this is a general problem with this cycle, we plotted in Figure \ref{SolarForecastingNeuralNetworksSumOverLatitudeCycle24} two forecasts, one with the data up to Cycle 22 (inclusive) and used it to forecast Cycle 24 (by concatenation): the green line, \textit{i.e.}\ exactly what we have done in Figure \ref{SolarForecastingNeuralNetworksSumOverLatitude}, and one where we used the entire dataset up to Cycle 23 (inclusive) to forecast Cycle 24: the orange line. The results seem to show an improvement in terms of a more accurate
cycle maximum amplitude (but not in terms of the actual cycle start). This improvement is due, presumably, because first, we used more data, and second, we do not attempt to forecast too ahead in time, a clear
known problem with chaotic systems, even if this is thought to be a weakly chaotic system. The forecast with more data also compares well with other results in the literature \citep[see][and references therein]{Pesnell_2016}. Our results are, of course, about sunspot areas. We can attempt to convert them into sunspot numbers by first, doing the same average or smoothing for the predicted sunspot areas (we shall call this $\bar{A}$) that is reported in the 13-month smoothed monthly sunspot number and, second, by noting the empirical relationship
 between sunspot areas (A)
and the sunspot numbers (R) in \cite{2006STIN...0620186W}: $A/R \approx 13.6\times 0.6$, where the $0.6$ factor comes from the 
recalibration of international sunspot numbers introduced in July 2015 \citep{Clette2016}. For the purpose of these calculations, we actually calculate an estimation of this empirical relationship ourselves, \textit{e.g.}\ by calculating the average $\bar{A}/R$ for Cycles 22 and 23, where $A$ is the sunspot area \textit{per} Carrington rotation number, $\bar{A}$ is the value of $A$ averaged or smoothed over 13 months, and $R$ is the 13-month smoothed monthly sunspot number. 
We obtain the value $\bar{A}/R\approx 9.40$ which compares well with the value from \cite{2006STIN...0620186W}, \textit{i.e.}\ $A/R \approx 13.6\times 0.6 \approx 8.16$.

This would imply, given our forecast of a Cycle 24 maximum 13-month smoothed sunspot area coverage of $\bar{A}_{24} = 1521.66 \pm 539.20$ (summed over latitudes), that  Cycle 24 would have a maximum sunspot number of $R_{24} = 161.86 \pm 57.36$. The actual maximum
according to the data in \cite{sidc} occurred around April 2014 with the value $R_{24}=116.4$, based on 13-month smoothed monthly sunspot number  data. Our value
is within the error  range, which is good, but our maximum shows up almost two years earlier than the real solar maximum. It is known that the Cycle 23 to 24 transition was an unusual low and long quiet period \citep{Lockwood_2012} and that without further data or external data it is probably not possible for the neural network method
to forecast that unusual late start.

\begin{figure}
\resizebox{.9\hsize}{!}{\includegraphics{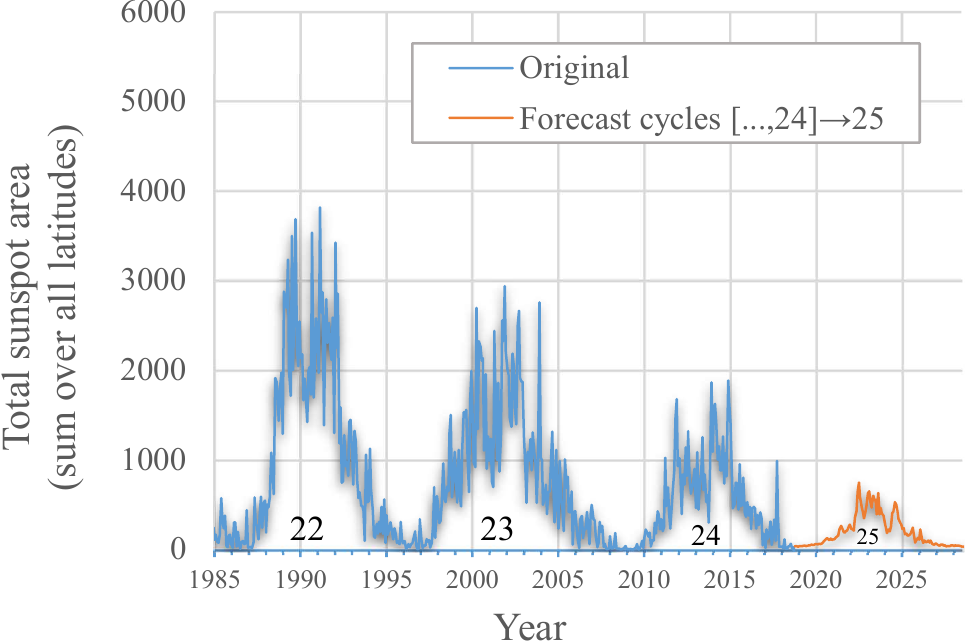}}
\caption{Forecast of the latitudinal sum of the sunspot area $A(t)$ for the next  Cycle 25 using a training set that includes all the data up to Cycle 24. We have marked the cycle numbers near the horizontal axis. It suggests the next cycle will be quite weak and peak around 2022--2023.}
\label{SolarForecastingNeuralNetworksSumOverLatitude25}
\end{figure}

Finally, before we conclude, we attempted to forecast the forthcoming Cycle 25, using all the spatial-temporal data we have. The results can be 
seen in Figure \ref{SolarForecastingNeuralNetworksSumOverLatitude25}. It seems to suggest that the next Cycle 25 will be around half of the intensity 
as the current one (Cycle 24), as measured by the sum of all sunspot areas (summed over the whole cycle and over all latitudes).  This would imply, given our forecast of a Cycle 25 maximum 13-month smoothed sunspot area coverage of $\bar{A}_{25}=538.09\pm 157.51$ (summed over latitudes), that the next Cycle 25 will have a maximum sunspot number of $R_{25}=57.24 \pm 16.76$. 
This suggests that this forthcoming Cycle 25 will be the weakest cycle in recorded history. As the saying goes, only time will tell. 
We also note that this value of $57.24$ is quite below the average\footnote{From a quick survey of 22 articles in the literature, we calculated an average of $R_{25}=106.03 \pm 34.77$ in terms of forecasts of the maximum for Cycle 25.} of values seen in the Cycle 25 prediction literature 
\citep[see][and references therein]{Hathaway2016a,Upton2018}.

Nonetheless, given our comments on the ``piano plot'' in the introduction, we do not want to put too much emphasis on this prediction, as we believe, based on that plot
and on the fact that the cycle amplitude is a metric that sums over all the detailed information on the spatial-temporal diagram, that one should
aim to compare the entire time and latitude forecast, and not just a numeric reduction of zero dimensionality such as the maximum sunspot number.

\section{Conclusion}
\label{conclusion}

We used the so-called feed-forward artificial neural network in an effort to forecast the solar sunspot butterfly diagram in space and time simultaneously. 
As far as we are aware, this is the first time this has been attempted using neural networks - 
all the work in the literature focus on using neural networks applied to pure sunspot temporal series, 
\textit{e.g.}\ the average sunspot number or the sunspot area coverage. To contrast, here we attempt to use neural networks to predict sunspots 
both in time and space.

The results show that the method can reproduce qualitatively some of the main features of the sunspot diagram, 
such as the overall cycle amplitude modulation, and the cycle sunspot migration to the equator.
However, there are limits to the forecast, namely the time horizon - the method does not seem to be able to forecast more than one cycle in advance, although this can be justified by the fact that the sunspot cycle is considered to be an example of a chaotic system, and therefore with a short predictability horizon. Also, the approach seems to be slightly biased towards medium cycles, the most common ones. This is clearly demonstrated in the analysis of the 
structure similarity index against the 
cycle strength. This is consistent on what was seen when using a
non-linear embedding method in \citep{covas2016}. In fact, there seems to be quite a parallel in the results of the two forecasting methods.  Further to that, we also predict, based on our method, that the upcoming Cycle 25 maximum will be around $R_{25}=57 \pm 17$. This implies that Cycle 25 will be the weakest cycle on record.

We also found that the spatial-temporal non-linear embedding method in \citep{2000PhRvL..84.1890P} points the way for the optimal
input layer representation for neural network forecasting, an empirical result that may show there is some general-purpose theorem waiting to be demonstrated. We plan to demonstrate this potential universality using other datasets in a forthcoming article.

We believe that the work points in the right direction, first that we ought to attempt to forecast in space as well as in time, and 
second that further improvements ought to be tried. The first improvement must surely
be to use recurrent networks, such as Elman networks \citep{COGS:COGS203} which are known to be more appropriate to model time series (although more complex to design and construct, hence why we started with simple and understandable feed-forward neural networks). The second improvement that we suggest is to incorporate related information as additional input(s), \textit{e.g.}\  to use solar magnetic field proxies such as the 10.7 cm radio flux and the 530.3 nm green coronal index \citep{2015SoPh..290.3095B}, and even geomagnetic proxies such as the {\it aa} indices \citep{1972JGR....77.6870M,1993GeoRL..20.2703N}, as these are also long time series with similar or higher
temporal frequency than the dataset we have used. 
In this article we wanted to focus on showing that neural networks can qualitatively model both the spatial and the temporal dynamics of the sunspot diagram and we plan to revisit the improvements mentioned above in future work.


Overall
we think forecasting in higher dimensions, particularly using neural networks and deep learning, even if it is harder computationally and more demanding in terms of the size of the data used, 
should point to other research possibilities within solar physics and within the emerging field of  space weather.





\begin{acks}
We would like to thank Reza Tavakol for very useful conversations regarding forecasting sunspots. We also would like to thank
Dr.\ David Hathaway for publishing the data that we used in this article. Finally, we would also like to thank the anonymous reviewer,
whose comments have helped to improve this article.
N.\ Peixinho acknowledges funding from the Portuguese FCT -- Foundation for Science and Technology (ref: SFRH/ BGCT/ 113686/ 2015).
CITEUC is funded by National Funds through FCT -- Foundation for Science and Technology (project: UID/ Multi/ 00611/ 2013) and FEDER -- European Regional Development Fund through COMPETE 2020 -- Operational Programme Competitiveness and Internationalisation (project: POCI-01-0145-FEDER-006922).
J.\ Fernandes acknowledges funding from the POCH and
Portuguese FCT – Foundation for Science and Technology (ref: SFRH/BSAB/143060/2018) and visiting facilities at Niels Bohr Institute (University
of Copenhagen).
\end{acks}

\disclaimer{The authors declare to have no conflicts of interest.}

\bibliographystyle{spr-mp-sola-eurico}
\bibliography{eurico_abbreviated} 

\end{article} 


\end{document}